\documentclass[useAMS,usenatbib]{mn2e}

\usepackage{txfonts,graphicx,epsfig,psfig,revsymb}

\def\be{\begin{equation}}
	\def\ee{\end{equation}}
\def\bea{\begin{eqnarray}}
\def\eea{\end{eqnarray}}

\def\lambdab{\lambdabar}

\def\mt {}
\def\no {}

\def\non {}

\def\gb {}
\def\fl {}
\def\gbn {}

\title[Stochastic wake field acceleration in GRB]{Stochastic wake field particle acceleration in Gamma-Ray Bursts}
\author[Barbiellini et al.]{G.Barbiellini$^{1}$, F.Longo$^{1}$, N.Omodei$^{2}$, A.Celotti$^{3}$ and M.Tavani$^{4, 5}$\\
$^{1}$Dipartimento di Fisica, Universit\`a di Trieste and INFN, Sezione di Trieste, via Valerio 2, 34100, Trieste, Italy\\
$^{2}$INFN, Sezione di Pisa, Edificio C - Polo Fibonacci - Largo B. Pontecorvo 3, Pisa, Italy\\
$^{3}$SISSA, via Beirut 2, 34100, Trieste, Italy\\
$^{4}$INAF-IASF, via Fosso del Cavaliere 100, Roma, Italy\\
$^{5}$Dipartimento di Fisica, via della Ricerca Scientifica, Universit\`a Tor Vergata, Roma (Italy)}

\begin{document}
\date{Received  / Accepted}

\maketitle

\begin{abstract}
{\mt
Gamma-Ray Burst (GRB) prompt emission can, for
specific conditions, be so powerful and short-pulsed to strongly
influence any surrounding plasma. In this paper, we briefly
discuss the possibility that a very intense initial burst of
radiation produced by GRBs satisfy the intensity and temporal
conditions to cause stochastic {\it wake-field particle
acceleration} in a surrounding plasma of moderate density. Recent
laboratory experiments clearly indicate that powerful laser beam
pulses of tens of femtosecond duration hitting on target plasmas
cause efficient particle acceleration and betatron radiation up
to tens of MeV . We consider a simple but realistic GRB model for
which particle wake-field acceleration can first be excited by a
very strong low-energy precursor, and then be effective in
producing the observed prompt X-ray and gamma-ray GRB emission.
We also briefly discuss some of the consequences of this novel GRB
emission mechanism.
} 

\end{abstract}

\begin{figure}%1
\includegraphics[width=0.8\hsize]{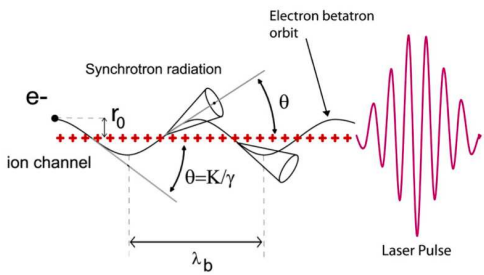}
\caption{Oscillation and radiation produced by a relativistic electron oscillating in an ion channel}
\label{fig::plasma}
\end{figure}

\section{Introduction}%1

{\mt 

Gamma-Ray Bursts (GRBs) are major explosive phenomena of our
Universe in need of an explanation.} The prompt GRB emission in
the hypothesis of beamed {\mt radiation} (with opening angle
$\theta_j = 0.1$) is  a very powerful gamma energy radiation with
typical emitted energy in the range of 10$^{51}$ erg lasting a
few tens of seconds and with a peak photon energy $E_p$ of a few
hundreds of keV. The compact nature of the GRB progenitors and
the observed photon energy spectra most likely require both
photon propagation transparency and relativistic motion of the
emitting source. The evolution of a typical GRB can be for simplicity divided in two
different phases: the first phase occurs close to the progenitor
and energizes and collimates the emitting particles.  This first
phase {\mt may be related to the} emission of quasi thermal or
Comptonized photons (Meszaros \& Rees, 2005). A second step can {\mt lead to
efficient particle acceleration far out from the central source,
and eventually to efficient radiation} by synchrotron emission in
internal {\mt and/or} external shocks.

{\mt In this paper, we investigate a model of GRB production and
evolution that assumes: (1) the existence of a thermal-like
precursor of very large intensity following the energy release
from relativistically moving shells or hydrodynamic fronts
originating from the GRB compact object; (2) the radiative
interaction of such a precursor (with a given power-spectrum as
seen by an observer at rest) with a surrounding (initially
not-ionized) gas shell at rest of density $n \sim 10^9 \rm
cm^{-3}$ at the approximate distance $R \sim 10^{15}\, $cm; (3)
propagation of the radiative front within the shell, with
possible activation of wake-plasma particle acceleration at the
outwardly propagating radiative shock; (4) wake-field particle
acceleration  producing a generation of non-thermal electrons
whose kinetic energy distribution function extends up to $ > 100
\,$MeV; (5) betatron radiation of these energized electrons
producing the observed non-thermal prompt GRB high-energy
emission in the X-ray/gamma-ray energy ranges.
{\gb The existence of an amount of material corresponding to a Thompson optical depth ~1} {\no is derived by the observation made by BATSE of } {\gb late bumps in the GRB light curve (Connaughton 2002), interpreted as a Compton reprocessed emission from  dense circumburst material at $R=10^{15}$~cm (Barbiellini et al., 2004).}
The physical origin of this material is at present subject of very intense simulations of the WR wind, where the wind could be compressed by the preceding supernova explosion. 
In this paper we would like to consider only the experimental evidence.
In this scenario, we assume that the energy attributed to the precursor (or to any electromagnetic signal entering the plasma) is {\gb $E_\gamma$/$\epsilon$ where $E_\gamma$ is the energy in the prompt emission and  $\epsilon$ is the conversion efficiency of the precursor energy in the prompt emitted radiation}. Since the photon component of the precursor has quasi-thermal energy, the absorption is very large; the highest energy component of the prompt is absorbed at a level of 70\%.}
Our GRB emission model does not imply a relevant role for the
intrinsic or MHD-driven comoving magnetic field in the ejecta
and/or radiative fronts. From this point of view, it provides an
emission scenario quite different from those assumed  in current
theoretical models of prompt GRB emission such as the synchrotron
shock model (Tavani 1996; Lloyd \& Petrosian 2002), the jitter
radiation model (Medvedev 2000), the diamagnetic relativistic
pulse accelerator (Liang et al. 2003), and the saturated
Comptonization model (e.g., Liang et al. 1997). 

{The very recent technical progress on high power laser beams
has {\mt provided} experimental evidence {\mt supporting} the
concept of wake-field {\mt particle} acceleration in a plasma
excited by an electromagnetic wave of duration comparable {\mt  or
shorter than } the inverse of the plasma frequency (Tajima and Dawson 1979).
A laser beam with a duration of 30 femtoseconds and a power of 50
TeraWatts, illuminating a gas jet 1 mm long with a density of
10$^{19}$ particles per cm$^3$ {\mt is able to accelerate}
electrons to an energy of few {\mt tens} of MeV. An X-ray beam of
{\mt a } few keV generated by the electron radiation loss in the
plasma {\mt was observed by Ta Phuoc et al. (2005) } with a characteristic
angular distribution {\mt resembling that of synchrotron
radiation } (this motivates the qualification used in the
literature of ``Laser driven synchrotron radiation"). 
Previous connections between the laboratory plasma physics and the astrophysical plasmas have been suggested by  Chen et al. (2002), which proposed a wake field acceleration excited by the Alfven shocks, producing Ultra High Energy Cosmic Rays (UHECR). Simulations using three dimensional particle-in-cell code, have been recently done by Ng \& Noble (2005) studying the dynamics of a neutral, narrow, electron-positron jet propagating through an unmagnetized electron-ion plasma.
In  {\mt this} paper, following the {\mt general} hypothesis already suggested by Tajima \& Dawson (1979), the processes of acceleration and radiative
losses induced by ultra short laser pulses on the gas target, {\mt validated by recent experimental data}, are {\mt suggested} to be the source of the GRB prompt emission. {\mt In our case, } the
laser role is played by {\mt GRB} precursor photons {\gb emitted by the $\gamma$ and $e^\pm$ shells collimated in an angle $\theta_j$ and moving with high relativistic velocity} and the plasma
density {\mt of the target gaseous nebula } is assumed to be on
the average $10^9$ cm$^{-3}$ over the region {\mt producing} the
prompt emission at a distance from the central source of about
$10^{15}$ cm.

{\mt In order to activate the {\gb coherent} wake-field particle acceleration in
the target plasma {\gb in the laboratory, where $n_{e0} \sim 10^{19}$ cm$^{-3}$, the threshold power surface density is 
$$
\sigma_{Wt} \sim 10^{18} {\mathrm{W cm}}^{-2}
$$
}}

{\gb Furthermore, assuming a coherent precursor emission} the
two conditions need to be satisfied: (1) the
radiation has to be suddenly delivered (in the form of pulses or
micro-bursts) to the target gas within ultra-short timescales
$\tau_p$ (that for our conditions need to satisfy the condition
$\tau_p < 10^{-9} \,$s in the observer/nebula frame); and (2) the
radiation intensity delivered within these ultra-short pulses has
to be sufficiently large to activate microscopic plasma electron-ion separation. 
In the following, we assume that the GRB initial fireball producing a very intense radiative precursor is satisfying condition (2). 
{\no In particular, this is the key point in the analogy GRB-Lab plasma acceleration is that the coherent radiation of laboratory lasers can be translated into non-coherent (stochastic) radiation from the precursor if a price is payed in terms of {\gb power density}.
The critical parameter in the coherent Lab plasma acceleration is the power  surface density threshold. We show below that $\sigma_{Wt}$ scales linearly with particle density. Its value is assumed from the experiment:

\begin{equation}
\sigma_{Wt} = 3\times 10^{18}(\frac{n}{10^{19}~cm^{-3}})\mbox{ W cm$^{-2}$}
\end{equation}
}
{\gb In the hypothesis that the GRB precursor carries an energy $E_{\mathrm pr} = E_\gamma/\epsilon$ and assuming $t_{\mathrm pr} \sim t_{grb} \sim 30 / (1+z)$ s the power density on the surface $R^{2}\theta^{2}=10^{27}$~cm$^2$ is:
\begin{equation}
\sigma_{Wgrb}=\frac{E_{\mathrm pr}}{t_{grb} R^2 \theta^2} \geq 2\times 10^{16} \mbox{ W cm$^{-2}$}
\end{equation}
which exceeds by several order of magnitude (around $10^8$) the power density threshold (scaled at $n=10^9$~cm$^{-3}$).

 The threshold to initiate stochastically the wakefield mechanism is equal to the coherent power density multiplied by the factor $\sqrt{t_{\mathrm pr}/\tau_{p}}$ that is is the probability to produce by chance the coherent WFA, independently in time and space. 
%by the square of the stochasticity parameter, proportional to $\sqrt{t_{\mathrm pr}/\tau_{p}}$. This stochastic factor is the probability to produce by chance the coherent WFA, independently in time and space. 

In other words, GRB not-coherent prompt emission is much more powerful than the coherent lab radiation: we assume that {\gb  the loss of coherence} is balanced by the enormous amount of power per square centimeter {\no as well as  the possibility of stochastic resonance: the wake field acceleration} threshold is overcome.

 Figure \ref{fig::plasma}
from Ta Phuoc et al. (2005) shows schematically the photon emission
principle. If the plasma is excited by the precursor photons, the
excited region and the emission of the radiated photons are {\mt
likely to be } in a cone of angular aperture equal to $\theta_j$
starting at a distance of $10^{15}$ cm from the central compact
object.

Stochastic acceleration and deceleration are in principle {\mt
related by plasma effects. However,} since they are statistically
independent, the two processes can be separated without loss of
generality. In the following, the acceleration phase is not
treated in detail. On the basis of the experimental results of Ta Phuoc et al. (2005) it is assumed that a power law energy spectrum of
electron and positron is produced. The accelerated beam enters in
the plasma region with constant gas density and radiates in the
plasma, following again the results of Ta Phuoc et al. (2005).

The value of the plasma density in the GRB environment is assumed
following Barbiellini et al. (2004). Because of the relation among the
precursor and the radiating electrons, to reproduce the GRB data a
stochastic treatment of the deterministic formulae of
Ta Phuoc et al. (2005) is applied.

In the laser-driven  {\mt betatron} radiation the emission follows the
scheme of Ta Phuoc et al. (2005)[fig.~1]. The angular distribution of the
emitted radiation is dominated by the wiggling angle of the
electron, $$\theta = \frac{K}{\gamma}$$ where $$K=\frac{\gamma
\omega_b}{c} r_0$$ {\mt with} $$\omega_b = \frac{\omega_p}{\sqrt{2 \gamma}}$$
the oscillating frequency, where $\omega_p=\sqrt{n e^2/(\epsilon_0
m_e)}$ is the plasma fundamental frequency and $r_0$ that could be estimated as:
$$
r_0=\frac{c}{\omega_p \sqrt\gamma}=\frac{\lambdabar_p}{\sqrt\gamma}$$
 is {\gbn equivalent to the} impact  parameter of the electron into the  positive plasma column
($e$ is the electron's charge). {\gbn Interpreting $\lambdabar_p = c/\omega_p$ as the screen distance of a charge inside the plasma, if the charge is moving relativistically $\lambdabar_b = c/\omega_b$ and $r_0$ represent respectively the longitudinal and transverse screening distance.} 

\section{Scaling relations in coherent plasma acceleration}

The key parameter in the coherent wakefield acceleration (CWFA) is the power surface density $\sigma_W$ needed to produce the complete charge separation over a lenght $\lambdabar_p$ on a timescale $\tau_p=\omega_p^{-1}$.

The energy volume density $\rho_\epsilon$ for complete separation could be espressed as:
$$
\rho_\epsilon = E^2 \epsilon_0 = \frac{\sigma^2}{\epsilon_0} = \frac{e^2n^2\lambdabar_p^2}{\epsilon_0}
$$
where $\sigma$ is the charge surface density. The quantity $\rho_\epsilon$ scales linearly with $n$ while the energy surface density $\sigma_\epsilon = \rho_\epsilon \lambdabar_p$ varies as:
$$
\sigma_\epsilon = \frac{e^2n^2\lambdabar_p^3}{\epsilon_0} \propto n^{1/2}.
$$
The key parameter $\sigma_W = \sigma_\epsilon \omega_p$ then scales as the plasma particle density $n$. 

%In the CWFA process also the equilibrium velocity of the accelerated electrons could be estimated as function of the plasma density. 
%The coherence condition requires that the oscillating electron with velocity $\beta$ remains in phase with the wave that generates the wakefield. This could be expressed by the following relation:
%$$
%M \left( -\frac{\lambdabar_b}{c} + \frac{\lambdabar_b}{\beta c \cos\theta} \right) = \frac{\lambdabar_b}{c} 
%$$
%$$
%(M - 1) \frac {\lambdabar_b}{\beta c}\left(\frac{1}{\cos\theta} - \beta \right) = 0 \\
%$$
%{\nom where M is an integer representing the number of radiation length after wich the electron is in phase with the plasma wave}. This holds if
%$$
%\left( 1 + \frac{\theta^2}{2} - \beta \right) = 0
%$$ 
%then $\gamma^2 \theta^2 = \mathrm{const}$. 

Using the relation among $\theta$ and the plasma quantities $\theta = \frac{r_0}{\lambda_b} \sim 1/\gamma $, since $r_0 = \lambdabar_p/\sqrt{\gamma}$ it is possible to derive:
\begin{equation}
\gamma r_0^2 \propto n^{-1}
\label{ron}
\end{equation}
%{\gbn as expected since $r_0 = \lambdabar_p/\sqrt{\gamma}$}. 

Let us consider an electron radiating with the typical formula: 
\begin{equation}
\frac{dE}{dt} =  -m~c^2 \frac{d\gamma}{dt} = \frac{e^2\gamma^4}{6\pi\epsilon_0c^3}|a_{\perp}|^2
\label{dEdt}
\end{equation} 

where $a_{\perp}$ is the transverse electron acceleration 
$$
a_{\perp} = r_0 \omega_b^2 = r_0 \frac{\omega_p^2}{2\gamma}.
$$

The energy loss by the electron in one oscillating period is:
$$
\frac{d\gamma}{dt}\propto \gamma^4 r_0^2 \omega_b^4
$$
with $\Delta t = \omega_b^{-1}$. At the equilibrium $\Delta \gamma \propto \gamma^{2.5} r_0^2 n^{3/2} = \mathrm{const}$. 
Recalling Eq.\ref{ron}, we obtain:
$$
\gamma^{3/2} n^{1/2} = \mathrm{const}
$$ 
and, hence, $\gamma \propto n^{-1/3}$. 
Furthermore, using again the relation $\gamma r_0^2 \propto n^{-1}$ it is also possible to derive that $r_0$ scales as $n^{-1/3}$.
Another way to obtain the previous relation is to consider that the restoring factor $K$ does not depend on the density $n$ so that:
$$
K\propto \omega_p~\gamma^{1/2}~r_0~\propto n^{1/2}~n^{-1/6}~r_0=\mathrm{const} 
$$
obtaining that $r_0 \propto n^{-1/3}$. 

Notice that the maximum electron Lorentz factor $\gamma_{\mathrm{max}}$ measured by Ta Phuoc et al. (2005) is, roughly $\gamma_{\mathrm{max}} \sim 200$. 
For a density of $10^9$~cm$^{-3}$, the electrons can be accelerated up to:
$$
\gamma_{\mathrm{max}} 
	\sim 200 \left(\frac{10^{19}}{10^9}
	\right)^{1/3} \sim 4.3\times 10^5
$$

In case of low densities, wake field acceleration can be as effective as producing high energy cosmic rays and, consequently gamma rays.

%$$
%the transverse oscillation from $\theta^2 \sim 2\times10^{-2}$ or $E_\mathrm{peak}\sim 2$ keV:
%$$
%r_0 = \frac{40}{n^{1/3}} \mathrm{cm}
%$$
%}

Following Ta Phuoc et al. (2005) and using the
fundamental frequency of radiation $\omega_f$ in the case the
plasma acts as a wiggler {\mt we have}
$$ \omega_f = \frac{3}{2} \gamma^3 c r_0
\frac{\omega_p^2}{2 \gamma c^2} = \frac{3}{4} c \gamma^2
\frac{r_0}{\lambdab_p^2} $$.

{\mt We obtain} an estimate of the peak energy of the emitted radiation $E_{\rm
peak}$: $$ E_{\rm peak} = \hbar \omega_f = \frac{3}{4} \hbar  c
\gamma^2 \frac{r_0}{\lambdab_p^2} $$

%The relevant quantities in the wakefield acceleration-emission process, $r_0$, $\lambdabar_p$ and $\gamma$ scale with the plasma density respectively as $n^{-1/3}$, $n^{-1/2}$ and $n^{-1/3}$. 

The energy distribution of the accelerated electrons is 
$$
N(\gamma) = N_0 \gamma^{-2}
$$
assuming the power law index again from the experiment (Ta Phuoc et al. 2005). 
%The mean peak energy of the radiated photons is in the $\sim$ keV energy band independently from the plasma density. 

{\fl The peak energy computed as an average on the electron spectrum is then:} 
$$
\langle E_{\rm{peak}} \rangle = \frac{3}{4} \hbar  c \frac{r_0}{\lambdabar_p^2}\gamma_{\rm{min}}\gamma_{\rm{max}}
$$

\section{GRB gamma emission in the stochastic wake field acceleration regime}%3

We address here the GRB prompt radiation produced by electrons radiating in a plasma excited by a very intense
precursor playing the role of a ``laser".
The basic formulae of laser-driven  {\mt betatron} radiation are assumed taking into account the
stochastic nature of the acceleration and the radiation phases.
Let us now apply these results to the case of a plasma accelerated
by the GRB precursor photons. We will treat the motion of the
relativistic electrons radiating in the plasma. The fundamental
difference with the above treatment resides in the stochastic
nature of acceleration and radiation in the case of the GRB
emission.

If the power surface density $\sigma_W$ at any position of the electrons producing the GRB exceeds the value measured by the experiment (Ta Phuoc et al. 2005) scaled at the GRB density and multiplied by the stochastic factor, the radiated photons of the GRB will be emitted from any accelerated electron provided that the stochastic angle of the electrons $\theta_{\rm{eff}}$ remain below the jet opening angle $\theta_j$. 
The elementary angle of the electron oscillation is:
$$
\theta_i = \frac{r_0}{\lambdabar_b}
$$
The stochastic motion of the electrons in the plasma results in an effective solid angle spanned by the electrons:
$$
\theta_{\rm{eff}}^2 = \left( \frac{R}{\lambdabar_b}\right) \theta_i^2
$$  
where we have introduced the adimensional quantity $N$:
$$
N = \left( \sqrt{\frac{R}{\lambdabar_b}} \right) = \sqrt{\frac{R}{\sqrt{2~\gamma}~\lambdabar_p}}
$$
where $R$ is the distance travelled by the electron inside the excited plasma. This will be used to estimate
the stochasticity of the electron geometrical path.

An electron with energy $E = \gamma mc^2$ wiggling inside the plasma with density $n$ will contribute to the GRB detectable emission for a path $R_{\theta}$ if:
\begin{equation}
[\theta_{\rm{eff}}(R_{\theta}, \gamma, n)]^2 \leq \theta_j^2
\label{theta:eq}
\end{equation}

Putting esplicitly the dependence on the path $R_{\theta}$ the previous equation could be written as:
$$
R_{\theta} \frac{r_0^2}{\lambdabar_p^3} (2\gamma)^{-1.5}  \leq \theta_j^2
$$

This relation expressed in terms of the plasma density and using the scaling laws for $r_0$, $\lambdabar_p$ and $\gamma$ derived previously could be written as:
  
%$$
%R_\theta(\gamma) =  \frac{2.8~\theta_j^2 \gamma^{1.5} \lambdabar_p^3}{r_0^2}\sim 7\times 10^6 cm \frac{\gamma^{3/2}\theta_j^2}{n^{5/6}}
%$$

$$
R_\theta(\gamma) =  2\sqrt{2} \lambdabar_p \theta_j^2 \gamma^{5/2}
$$

{\fl The equation \ref{theta:eq} introduces a threshold for the electron emission over a distance $R$. The threshold Lorentz factor $\gamma_t$ is obtained requiring that even in the case of a large energy loss, the electron trajectory still remains within the energized cone. 

The effective energy loss in the stochastic acceleration regime is:
$$
\frac{d \gamma}{dt} = - \frac{2}{3} r_e r_0^2 \gamma^4 \frac{\omega_b^4}{c^3}\left( \frac{R}{\lambdabar}  \right)^{0.5}
$$
where for simplicity we have introduced the classical electron radius $r_e$
$$
r_e = \frac{e^2}{4\pi \epsilon_0 m_e c^2}
$$ 
and where we have multiplied the coherent energy loss by the stochastic factor $\sqrt{ R/\lambdabar}$. This factor is our best estimation of the effect of the turbulence of the circumburst medium. Expressing the previous relation in terms of $\gamma$,  and substituting $dt=c~dR$, we obtain:
%$$
%d \gamma \gamma^{-7/4} = 0.1 r_e r_0^2 \lambdabar_b^{-9/2} dR^{-3/2}
%$$

$$
d \gamma \gamma^{-3/4} = \frac{1}{11} r_e \lambdabar_p^{-5/2} dR^{3/2}
$$

$$
\gamma^{1/4} = \frac{r_e}{8.3} \lambdabar_p^{-5/2} R^{3/2}
$$
From this relation we estimate a typical ``energy loss'' distance $R_e$ equivalent to the path where the electron looses half of its energy. Integrating the previous equation  we get:%from $\gamma$ to $\gamma/2$ DA RIVEDERE ..
$$
%R_e \sim \frac{3.5 \lambdabar_p^3 \gamma^{-1/2}}{\left( r_e r_0^2\right)^{2/3}}
R_e \sim 4.1~\lambdabar_p\left(\frac{\lambdabar_p}{r_e}\right)^{2/3}\gamma^{1/6}
$$
and
$$
%\frac{R_e}{\lambdabar_b}= \frac{2.5\lambdabar_p^2}{\left(r_e r_0^2 \right) ^{2/3}}\gamma^{-1}
\frac{R_e}{\lambdabar_b}= 2.8~\gamma^{-1/3}\left(\frac{\lambdabar_p}{r_e}\right)^{2/3}
$$
%\frac{4~(2-2^{1/4})}{3}\gamma^{-3/4} = \frac{R_e}{6}\lambdabar_p^{-9/2}r_e~r_0^2~(c t_{pr})^{0.5}$$

%or, obtaining $R_e$ as a function of the initial lorenz factor of the electron:
%$$
%R_e (\gamma) = \frac{6.49 \gamma^{-3/4} \lambdabar_p^{9/2}}{r_e r_0^2 (c t_{pr})^{0.5}}\sim \frac{3.672\times 10^{30}cm}{n^{19/12}\sqrt{t_{pr}}}
%\gamma^{3/4}
%$$

The value of $\gamma_t$ is then obtained by requiring 
$$
R_{\theta}(\gamma_t) = R_e (\gamma_t)
$$
obtaining 
$$
%\gamma_t  = \frac{1.62 \times 10^7}{n^{1/3}~t_{pr}^{2/9}\theta_j^{8/9}}
%\gamma_t  =1.1\left(\frac{r_0}{r_e}\right)^{1/3} \theta_j^{-1}
\gamma_t  =1.2\left(\frac{\lambdabar_p}{r_e}\right)^{2/7} \theta_j^{-6/7}
$$
The opening angle of the jet $\theta_j$ imposes a threshold on the energy of the emitting electron wiggling with an effective angle so that the mean value of the peak energy has to be computed over the electron spectrum integrated only between $\gamma_t$ and $\gamma_{\rm{max}}$.

{\fl Taking into account the stochastic nature of the phenomenum due to the external density turbolence, we can calculate the predicted value for  $\langle E_{\rm{peak}}\rangle_{\rm{eff}}$ in the stochastic regime, simply multiplying the coherent value of $\langle E_{\rm{peak}}\rangle$ by the stochastic factor $\sqrt{R_e/\lambdabar_b}$. After an integration between $2\gamma_t$ and $\gamma_t$, assuming that only these electrons contribute to the observed radiation, we obtain:

% (notice that $\langle E_{\rm{peak}}\rangle\propto \lambdabar_p^{-2}\propto n$. The density scales, as usual, with the stochastic factor N):
 %$$
%\langle E_{\rm{peak}}\rangle_{\rm{eff}} = \frac{3}{4} \hbar  c \frac{r_0}{\lambdabar_p^2}\gamma_t\gamma_{\rm{max}} \sqrt{\frac{R_e}{\lambdabar_b}}
%$$
%$$
%\langle E_{\rm{peak}}\rangle_{\rm{eff}} = 9 \hbar c\left(\frac{r_0}{r_e}\right)^{1/3} \frac{\gamma_t^{1.5}}{\lambdabar_p} 
%$$

$$
%\langle E_{\rm{peak}}\rangle_{\rm{eff}} =  \frac{1.8 \times 10^{-7}}{\lambdabar_p} \left( \frac{r_0}{r_e} \right)^{1/3} \gamma_t^{1.5} \rm{keV} 
\langle E_{\rm{peak}}\rangle_{\rm{eff}} = 2.5 \frac{\hbar c}{\lambdabar_p}\left(\frac{\lambdabar_p}{r_e}\right)^{1/3}\gamma_t^{4/3}
$$
$$
%\langle E_{\rm{peak}}\rangle_{\rm{eff}} =  \frac{2.0 \times 10^{-7}}{\lambdabar_p} \left( \frac{r_0}{r_e} \right)^{5/6} \theta_j^{-1.5} \rm{keV} 
\langle E_{\rm{peak}}\rangle_{\rm{eff}} =  3.0 \frac{\hbar c}{\lambdabar_p} \left(\frac{\lambdabar_p}{r_e}\right)^{5/7}\theta_j^{-8/7}
$$
%Expressing the previous equation in terms of observable quantities such as plasma density and jet opening angle  we predict a value of  $E_{\rm{peak}}$:
%$$
%\langle E_{\rm{peak}}\rangle_{\rm{eff}} = \frac{5.45 \times 10^5~\rm{keV}}{\theta_j^{1/3}~t_{pr}^{1/3}~n^{1/3}}
%$$
{\non From the previous relations, we have:}
$$
\langle E_{\rm{peak}}\rangle_{\rm{eff}} \approx 25~\rm{keV}\frac{n}{10^9}\theta_j^{-8/7}
$$

Assuming the density $n = 10^9$ cm$^{-3}$ obtained by the analysis of the Compton tail phenomenum (Barbiellini et al. 2004), and {\non $\theta_j$=0.1 (\gbn $\theta_j \approx 6^{\circ}$)} we  obtain in the GRB source frame:

$$
\langle E_{\rm{peak}}\rangle_{\rm{eff}} \approx 350~\rm{keV}
$$
%assuming $r_0(n=10^{19})=0.3 \mu\rm{m}$ as suggested by Ta Phuoc et al. (2005) and using the scaling law $r_0 \propto n^{-1/3}$.  
}

\section{Spectral - Energy correlations}

{\fl Noticing that the quantity $\gamma_t$ introduces then a dependence of $\langle E_{\rm{peak}}\rangle$ on $\theta_j$, we use} the above derived formulae for the peak energy of the emission and the effective angle of the radiation to derive an estimate for the recently obtained Amati (Amati et al., 2002) and Ghirlanda (Nava et al. 2006) relations.

{\fl Considering the electrons with an energy spectrum similar to that derived experimentally in the coherence regime,} with Lorentz factor between $\gamma_t$ and $2\gamma_t$, the mean useful electron energy is:
$$
E_\gamma = N_0 m c^2 \frac {\int \gamma \gamma^{-2} d\gamma}{\int \gamma^{-2}d\gamma} = 2 \ln{2} N_0 mc^2 \gamma_t 
$$

If each electron of energy greater than $mc^2\gamma_t$ emit $n_\gamma$ of mean energy $\langle E_{\rm{peak}}\rangle$ then the energy conservations requires:

$$
mc^2 \langle \gamma \rangle = n_\gamma \langle E_{\rm{peak}}\rangle
$$
$$
% (2 \ln{2}) mc^2 \gamma_t = n_\gamma \times 9 \hbar c\left(\frac{r_0}{r_e}\right)^{1/3} \frac{\gamma_t^{1.5}}{\lambdabar_p}
 (2 \ln{2})~mc^2 \gamma_t = n_\gamma \frac{\hbar c}{\lambdabar_p} \left(\frac{\lambdabar_p}{r_e}\right)^{1/3}\gamma_t^{4/3}
$$

%$$
% (2.2 \ln{2}) mc^2 \left(\frac{r_0}{r_e}\right)^{1/3} \theta_j^{-1} = n_\gamma \times  \frac{2.0 \times 10^{-7}}{\lambdabar_p} \left( \frac{r_0}{r_e} \right)^{5/6} \theta_j^{-1.5}
%$$

{\fl  
The energy conservation requires {\gbn $n_\gamma \propto \gamma_{t}^{-1/3}\propto\langle E_{\rm{peak}}\rangle^{-1/4} $}. The total energy $ E_\gamma$ being composed of the emission by the $N_0$ electrons results}
$$
N_0 mc^2 \langle \gamma \rangle = E_\gamma = N_0 n_\gamma \langle E_{\rm{peak}}\rangle =  N_0 n_\gamma \langle E_{\rm{peak}}\rangle^{1/4} \langle E_{\rm{peak}}\rangle^{3/4}
$$
%$$
%E_\gamma = N_0  n_\gamma \times \left[ \frac{2.0 \times 10^{-7}}{\lambdabar_p} \left( \frac{r_0}{r_e} \right)^{5/6} \theta_j^{-3/2} \right]^{1/3} \times \left[ \frac{2.0 \times 10^{-7}}{\lambdabar_p} \left( \frac{r_0}{r_e} \right)^{5/6} \theta_j^{-3/2} \right]^{2/3} 
%$$
Obtaining:
$$
E_\gamma =  N_0 \times \rm{const}\times \langle E_{\rm{peak}}\rangle^{3/4}
$$

If the isotropic energy $E_{\rm{iso}}$ is introduced:
$$
E_{\rm{iso}} \approx \frac{E_\gamma}{\theta_j^2} = \frac{N_0 n_\gamma \langle E_{\rm{peak}}\rangle}{\theta_j^2} \propto \theta_j^{-2} \langle E_{\rm{peak}}\rangle^{3/4} \propto  \langle E_{\rm{peak}}\rangle^{5/2}
$$
Inverting the previous equation we obtain $E_{\rm {peak}}\propto E_{\rm{iso}}^{0.4}$ as outlined in the so called ``Amati'' relation.
%being $\theta_j^2 \propto \langle E_{\rm{peak}}\rangle^{-2.}$, obtaining the ``Amati'' relation.}

\section{Conclusion}

{\mt We studied the properties of a novel mechanism for prompt
GRB radiation based on plasma laser-wake acceleration and
radiation induced by an intense precursor irradiating a static
gas nebula.} The stochastic radiative emission of the energized
plasma produces a photon emission with angular distribution
dominated by the electron trajectories. {\mt Under general
conditions applicable to our GRB model (precursor, static cloud,
laser-wake acceleration conditions) we showed that there is a
relation between the observed jet property ($\theta_j$) and an
electron energy threshold. In our mechanism the energy of the
emitting wiggling electrons is the typical energy of the
laser-wake betatron radiation.  Furthermore, we derived a relation
between source properties such as the collimation angle and the
emitted energy. }

\section*{Acknowledgements}
We would like to thank Ugo Amaldi for the suggestion to investigate the plasma wake field acceleration as an efficient mechansim, Victor Malka and Danilo Giuletti for the illustration of the fundamental elements on plasma acceleration. One of us (G.B.) would like also to express gratitude to Giuseppe Cocconi for the wise suggestion on how to deal with the comparison of events taking place within environments with scale different by many order of magnitude.

\end{document}